# Magnetic-field-induced ordered phase in the chloro-bridged copper(II) dimer system [Cu$_2$(apyhist)$_2$Cl$_2$](ClO$_4$)$_2$

R. S. Freitas[1], W. A. Alves[2] and A. Paduan-Filho[1]

[1]Instituto de Física, Universidade de São Paulo, 05314-970 São Paulo, SP, Brazil.
[2]Centro de Ciências Naturais e Humanas, Universidade Federal do ABC, 09210-580 Santo André, SP, Brazil.

*Corresponding author: freitas@if.usp.br

## ABSTRACT

Specific heat and magnetization measurements of the compound [Cu$_2$(apyhist)$_2$Cl$_2$](ClO$_4$)$_2$, where apyhist = (4-imidazolyl)ethylene-2-amino-1-ethylpyridine), were used to identify a magnetic-field-induced long-range antiferromagnetic ordered phase at low temperatures (T < 0.36 K) and magnetic fields (1.6 T < H < 5.3 T). This system consists of a Schiff base copper(II) complex, containing chloro-bridges between adjacent copper ions in a dinuclear arrangement, with an antiferromagnetic intradimer interaction $|J_{intra}|/k_B \approx 3.65$ K linked by an antiferromagnetic coupling $|J_{inter}|z/k_B \approx 2.7$ K. The magnetic-field-induced ordering behavior was analyzed using the mean field approximation and Monte Carlo simulation results. The obtained physical properties of the system are consistent with the description of the ordered phase as a Bose-Einstein Condensation (BEC) of magnetic excitations. We present the phase diagram of this compound, which shows one of the lowest critical magnetic field among all known members of the family of BEC quantum magnets.

# I. INTRODUCTION

An important area of research in magnetism is the study of systems with molecular units that are capable of manipulation, leading to the realization of compounds with different defined properties and providing insight into the realization of new quantum states of matter [1,2]. Since quantum effects are enhanced for spin-1/2 systems, there has been much theoretical discussion and experimental investigation addressing the magnetism of systems with low-spin values [3,4]. Several low-dimensional compounds have provided an excellent illustration of the magnetic-field-induced quantum critical point (QCP) in dimer systems with antiferromagnetic interactions with 3D-ordered phases [5]. Theoretically predicted some years ago [6], the Bose-Einstein condensation (BEC) in magnetic systems was first observed in the spin-½ dimer compound $TlCuCl_3$, in which the intra-copper interactions, $J_{intra}$, are antiferromagnetic [7,8,9]. This compound and the more recently studied $BaCuSi_2O_6$ system [10,11] are understood in the BEC scenario of weakly coupled spin dimers. The energy spectrum of these gapped system is a singlet ground state ($S=0$) separated by the exchange energy to an excited triplet ($S=1$). The application of a magnetic field closes this gap, lowering the triplet-state energy, which becomes degenerate with the singlet ground state. At this point, when the gap is closed, it is convenient to consider the triplet as bosonic quasi-particles. The possible existence of a weak interdimer interaction spreads the region where the system becomes gapless, leading to a long-range order described as a condensed triplet [12,13]. Figure 1 schematically shows this level crossing where the gapless phase is limited by the critical fields $H_{C1}$ and $H_{C2}$.

More recently, several other compounds showing quantum-phase transitions have been found. Besides copper dimers, these systems include another class of 1D materials composed by Ni ions [14]. Here the single ion anisotropy splits the $S = 1$ triplet of each $Ni^{2+}$ into a ground state and an excited doublet. In this configuration, in which the ground state is S=0, a small additional interaction is not enough to induce long-range order in zero magnetic field. One example of this behavior is observed in the compound DTN, $NiCl_2.4SC(NH_2)_2$, one of the most studied quantum spin system presenting a magnetic-field-induced 3D-ordered phase, described as a BEC [5,14]. A necessary condition for the use of the BEC formalism in these systems is the absence of anisotropy violating the rotational symmetry of the magnetic ions. Such analysis describing the magnetic ordered phase as a field-induced bosonic state have been quite successful providing information on elementary excitation, and explaining several properties such as magnetization, specific heat, phase boundaries, NMR relaxation and the Bose-glass phase [5,15, 16].

Preliminary magnetic investigation of the compound $[Cu_2(apyhist)_2Cl_2](ClO_4)_2$ indicates that it has a chemical and magnetic structure composed of dimers [17]. In this structure, the two magnetic ions of Cu are imbedded within a non-magnetic ligand group, interacting mainly via antiferromagnetic super-exchange Heisenberg coupling. The intradimer interaction, $J_{intra}$, was determined as $|J_{intra}|/k_B \sim 2$ K using the Bleaney-Bowers model for the magnetic susceptibility, and a mean-field approximation was used to estimate the exchange coupling $|J_{inter}|z/k_B = 1.3$ K among dimers [17]. Since the signal of the intra-dimer coupling is antiferromagnetic, the magnetic ground state of this compound was identified as a spin singlet with an excitation gap.

Here, we study the thermal and magnetic properties of $[Cu_2(apyhist)_2Cl_2](ClO_4)_2$ at ultra-low temperatures. Our results show the presence of a magnetic-field-induced magnetic long-range order, suggesting that this material is a new quantum magnet candidate that displays Bose-Einstein condensation of magnons.

## II. EXPERIMENTS

Powdered samples of $[Cu_2(apyhist)_2Cl_2](ClO_4)_2$, where apyhist = (4-imidazolyl)ethylene-2-amino-1-ethylpyridine, were synthesized using the procedure described elsewhere [17]. This binuclear compound crystallizes in a triclinic system with space group $P\bar{1}$. The X-ray structure determination revealed ionic structures consisting of one complex cation $[Cu_2Cl_2(C_{12}H_{14}N_4)_2]^{2+}$ and two respective perchlorate anions between dimer units [18]. The packing arrangement of the complex salt is shown in Fig. 2. The subsequent layers in the lattice are joined together by means of electrostatic forces between the oxygen atoms of the perchlorate groups and the N-H and C-H groups of the complex cations [18]. The subunits are held together principally through their bridging chloride ligands. Each copper atom is five-coordinate, and the bond angles at the copper center indicate that it adopts a distorted square-pyramidal geometry. The coordinated atoms are three donor atoms from the ligand *apyhist*, a chlorine atom (Cl) form the square base, and another chlorine atom occupying the fifth apical position, with Cu-Cl distances of 2.271 and 2.737 Å and a Cu–Cl–Cu angle of 87.46°. The Cu···Cu distance in the $[Cu_2(apyhist)_2Cl_2]^{2+}$ core is 3.478(1) Å [15]. The value of $\phi/R$, where $\phi$ is the Cu-Cl-Cu angle (°) and *R* is the longer Cu-Cl distance (Å), is 31.95 for [Cu$_2$(apyhist)$_2$Cl$_2$](ClO$_4$)$_2$ and is comparable to the values of 30.76 and 31.50 reported for the antiferromagnetically coupled dimers [Cu$_2$(*N,N,N´,N´*-tetramethylethylenediamine)$_2$Cl$_4$] and [Cu$_2$(*N,N*-dimethylenediamine)$_2$Cl$_4$], respectively [15]. For this *ϕ/R* value, Hodgson´s empirical correlation [19] predicts antiferromagnetic exchange in [Cu$_2$(apyhist)$_2$Cl$_2$](ClO$_4$)$_2$ with an exchange energy 2*J* of *ca.* -4.3 K.

The magnetic susceptibility (M/H) was measured using a SQUID magnetometer (Quantum Design, MPMS) at high temperatures (*T* > 2 K) and at low temperatures (down to 0.6 K) with a vibrating sample magnetometer adapted for use in a $^3$He cryostat. The specific heat data were obtained using a Quantum Design Dynacool system, equipped with a dilution refrigerator option, using a standard semi-adiabatic heat pulse technique under magnetic fields up to 9 T and temperatures down to 0.1 K. The addendum heat capacity was measured separately and subtracted. Measurements of the specific heat as a function of the applied magnetic field in a nearly constant temperature were obtained using very small heat pulses, resulting in a temperature change of the sample of less than 0.04 K during the measurements.

## III. ANALYSIS AND DISCUSSION

The temperature dependence of the magnetic susceptibility, *χ* = *M/H*, measured with *H* = 500 Oe is shown in Fig. 3. The use of a low value for the magnetic field is a necessary condition to guarantee the linear behavior of the measured magnetic moment with the field. The susceptibility increases with decreasing temperature until a rounded maximum is reached at *T* ~ 2 K followed by a susceptibility decrease, which is indicative of a non-magnetic spin singlet ground state. The data in the present work was measured down to *T* = 0.6 *K*, extending the temperature range of [17] to better estimate the magnetic coupling in the sample. To describe the magnetic behavior of this binuclear system we used the Heisenberg exchange Hamiltonian:

$$H = -J_{intra}S_1S_2 + g\mu_B S'H \qquad (1)$$

where the Zeeman term is added, *H* is the external magnetic field and *S´* is the total spin operator of the dimer with spins $S_1$ and $S_2$. To analyze the data of Fig.3 we used the numerical calculation of Johnston *et al.* for *S*=1/2 isolated dimers [20]:

$$\chi^* = \left[\frac{Ng^2\mu_B^2}{4kT}\right] \times \left[e^{\left(\frac{-J_{intra}}{kT}\right)}\right] \times \frac{\sum_{n=1}^{5}\frac{N_n}{t^n}}{\sum_{m=1}^{5}\frac{D_m}{t^m}} \qquad (2)$$

Here the parameters were $N_1$=0.634298982, $N_2$=0.1877696166, $N_3$=0.03360361730, $N_4$=0.003861106893, $N_5$=0.0002733142974, $D_1$=-0.1157201018, $D_2$=0.08705969205, $D_3$=0.0056313666688, $D_4$=0.0011040886574, and $D_5$=0.00006832857434. To take in account the interaction among dimers an interdimer interaction $J_{inter}$ should be incorporated to the Hamiltonian. Treating this interaction in the mean field approximation (MFA) the effective susceptibility $\chi$ at low field becomes [17]:

$$\chi = \frac{\chi^*}{(1-\gamma\chi^*)}, \qquad (3)$$

where $\chi^*$ is the susceptibility of an isolated dimer, $\gamma$ is a mean field correction given by $\gamma = J_{inter} z/Ng^2\mu_B^2$ and $z$ is the number of neighboring dimers. Equation (3) fits very well to the experimental data in Fig. 3 with the following parameters: $g = 2.1$, $|J_{intra}|/k_B = 3.7$ K, $|J_{inter}z|/k_B = 2.7$ K, both interactions being antiferromagnetic. These results are in good agreement with the ones based on the Bleaney-Bowers equation [17], in which the $J$ value is defined as half the one used in Eq. (1). The close agreement between the fit and the data below the maximum in the susceptibility curve indicates that at low magnetic fields the mean field approach for the interdimer interactions captures the fundamental physics of the problem that defines the thermodynamic properties of the system. The $g$ value is comparable with that obtained from earlier EPR measurements ($g = 2.14$) and from the value inferred from the saturation of the magnetization, $g = 2.03$ [17].

Even though the magnetic results at low field could be satisfactorily accounted for by the mean field approach, without showing any long-range order, the application of an applied magnetic field may drive the system to a more complex magnetic structure. As anticipated by the energy levels in the scheme shown in Fig. 1, a magnetic field may act to create a degeneracy between the ground state and the lowest excited singlet resulting in a quantum phase transition from a disordered paramagnetic to an induced long-range ordered phase. This ordered phase is expected by the analysis of the exchange parameters obtained from the zero-field magnetic data.

The temperature dependence of the specific heat $C(T)$ is shown in Fig. 4 for some selected magnetic fields. The first point to notice in the measurement under zero magnetic field is the absence of a sharp specific heat peak associated with a long-range ordered phase transition. This fact is an additional confirmation of the gapped singlet ground state of this compound at $H = 0$. A broad maximum centered at $T_D \approx 1.5$ K is clearly observed in the data measured with low applied magnetic fields ($H < 3.5$T). This feature is characteristic of short-range interactions and we associate it to the onset of dimer formation. With the increase of the applied magnetic field, this maximum at $T_D$ is suppressed due to the closing of the gap between the ground state and the first excited triplet. The curve measured under $H = 2$ T shows another rounded maximum at $T_N = 0.27$ K. This second maximum has a non-monotonic behavior as the applied magnetic field increases. In the range from $H = 2$ T to 3.5 T this maximum becomes sharper and shifts to higher temperatures. However, for $H > 3.5$ T it reverses its behavior and gets suppressed to lower temperatures. We associate this second anomaly to a long-range 3D antiferromagnetic order driven by the interdimer coupling $J_{inter}$. Similar field-induced magnetic ordering at low temperatures have been observed in others spin dimer compounds [5]. Finally, the increase of $C(T)$ at very low temperature (below ~0.2 K) may be ascribed to a magnetic nuclear contribution to the specific heat [21].

The electronic magnetic contribution to the specific heat $C_e(T)$ can be obtained by subtracting out the lattice $C_l(T)$ and nuclear $C_n(T)$ components from the total measured specific heat $C_e(T) = C(T) - C_l(T) - C_n(T)$. The lattice contributions can be estimated by fitting the high-temperature part of the total specific heat to an asymptotic series of odd powers of the temperature, which correspond to a low-frequency expansion of the Debye function $C_l(T) = aT^3 + bT^5 + cT^7$ [Ref. 22]. The nuclear component has the usual [23] temperature dependence $C_n(T) = \beta T^{-2}$ below $T \sim 0.2$ K, as shown in Fig. 4. This nuclear contribution has an applied magnetic field dependence and saturates for H > 2 T. The inset of Fig. 5 shows the subtracted magnetic electronic specific heat data measured at zero magnetic field. This result can be adjusted using the calculated specific heat for a Heisenberg $S = ½$ dimer [24]:

$$C_e(T) = 12R \left(\frac{J_{intra}}{2k_BT}\right)^2 \frac{e^{\frac{J_{intra}}{k_BT}}}{\left(1+3e^{\frac{J_{intra}}{k_BT}}\right)^2} \qquad (4)$$

This expression for isolated dimers gives a good description of the data and the corresponding antiferromagnetic exchange parameter, $|J_{intra}|/k_B = 3.62$ K, is in excellent agreement with the one obtained from the susceptibility data. The small discrepancies between the data and the fit may be attributed to the interdimer exchange coupling [25] and uncertainties in the lattice and nuclear contributions to the specific heat. The entropy change associated with the electronic magnetic degrees of freedom can be estimated by integrating the corresponding specific heat $\Delta S_e = \int \frac{C_e}{T} dT$. The result is shown in Fig. 5 and nicely meets the expected value for a $S = ½$ system, $\Delta S_e = R \ln 2$. The field-induced antiferromagnetic nature of the sample's ground state is confirmed by the temperature dependence of the electronic specific heat at low temperatures. The inset of Fig. 6 shows the total measured specific heat under H = 3.5 T and the electronic contribution $C_e$, obtained after the subtraction of the nuclear part, shows the characteristic $T^3$ behavior expected for antiferromagnetic magnons as displayed in the main panel of Fig. 6.

Figure 7 shows the magnetic field dependence of the specific heat measured at some selected temperatures. Two clear peaks can be seen at $H_{c1}$ and $H_{c2}$ for each temperature. These anomalies are related to phase transitions, marking the boundary of the field-induced ordered phase at $H_{c1}$ and the fully spin-polarized phase above $H_{c2}$. The asymmetry of the peaks closely resembles what is observed in the foremost BEC compound NiCl$_2$-4SC(NH$_2$)$_2$ [26], where $C_{c2}/C_{c1} \approx 6$. In our case we observe a smaller ratio value, ~1.6, which is probably related to the polycrystalline nature of the sample and the higher temperature of our measurements when compared to the ones in Ref. 26. The use of specific heat measurements to identify the critical fields at the lowest temperatures (below ~0.2 K) in our compound is hindered by the fact that the copper nuclear contribution dominates the measured specific heat as the temperature decreases.

The experimental results from the anomalies of the specific heat measurements are summarized in Fig. 8. The region inside the boundary determined by $H_{c1}$ (T) and $H_{c2}$ (T) corresponds to the field-induced long-range ordered phase. It is noteworthy to point the asymmetry of the phase diagram dome, a general characteristic observed for other quantum magnets with BEC phases [5]. This 3D-ordered phase has been discussed using a spin-pair model with mean field approach by Tachiki and Yamada [27] to explain the phase diagram of Cu(NO$_3$)$_2$·2.5H$_2$O and, more recently, by Nohadani *et al.* [28] using Quantum Monte Carlo (QMC) simulations in the BEC scenario of magnons. We will use these two approaches to correlate the magnetic coupling parameters of our compound with the critical fields in the phase diagram of Fig. 8. In the first approach the equations to determine the critical fields at zero and maximum temperatures of the phase diagram are given by [27]:

$$g\mu_B H_{c1}^{MFA}(0) = J_{intra} - \frac{J_{inter}z}{2} \quad (5)$$

$$g\mu_B H_{c2}^{MFA}(0) = J_{intra} + J_{inter}z \quad (6)$$

$$T_{max}^{MFA} = \frac{J_{inter}z}{4k_B} \quad (7)$$

Using the magnetic coupling values obtained in the susceptibility and specific analyses in the set of equations above gives $H_{c1}^{MFA}(0) = 1.66$ T, $H_{c2}^{MFA}(0) = 4.53$ T and $T_{max}^{MFA} = 0.67$ K, where we have assumed z = 6. On the other hand, our magnetic coupling values can be used to extrapolate another set of critical fields using the *QMC* results (Figs. 1 and 3 of Ref. 28) yielding $H_{c1}^{QMC}(0) = 1.69$ T, $H_{c2}^{QMC}(0) = 3.93$ T and $T_{max}^{QMC} = 0.385$ K. Both sets of critical fields are in reasonably good agreement with the visually extrapolated values obtained from the phase diagram in Fig. 8, $H_{c1}$ (0) = 1.6 T, $H_{c2}$ (0) = 5.3 T and $T_{max}$ = 0.36 K. The discrepancies between the calculated and determined upper critical field values, $H_{c2}$ (0), may be explained by the uncertainties in the extrapolated phase boundary to zero temperature and the excess of fluctuations near this transition field, as reflected in the higher value of the specific heat peak relative to the lower transition field $H_{c1}$ (0) [26,29]. The value of the maximum temperature in the phase diagram dome in Fig. 8 is however less susceptible to extrapolation uncertainties. This quantity is more consistent with the QMC results for a BEC of magnons than the one predicted using the mean field approximation of a regular field-induced phase transition in a spin-pair model.

## IV. SUMMARY AND CONCLUSIONS

In summary, our magnetic and thermal experiments on [Cu$_2$(apyhist)$_2$Cl$_2$](ClO$_4$)$_2$ show that this compound remains in a gapped phase with no long-range order under zero magnetic field in the whole investigated temperature range. The zero-field low-temperature behavior can be reproduced by different dimer models with antiferromagnetic intradimer |$J_{intra}$|/$k_B$ = 3.75 K and interdimer |$J_{inter}$|z/$k_B$ = 2.7 K interactions. Specific heat measurements reveal that the gap is closed by the application of a magnetic field, leading to an induced antiferromagnetic ordered phase at low temperatures. The phase diagram of this compound is presented and its dome shape, determined by the critical fields at zero and maximum temperatures, is consistent with the ones expected for a Bose-Einstein condensation of magnons. We hope our results will stimulate other studies with additional probing tools and the use of single crystalline samples to fully characterize the nature of the field-induced phase in this material. The confirmation of a Bose-Einstein condensation of magnons in this compound puts it in a special position, as it possesses the record lowest critical magnetic fields of all known members of the BEC family of quantum magnets. These small critical field values would allow a fully investigation of the exotic disordered-induced BEC phase recently proposed to exist in DTN [30,31].

## ACKNOWLEDGMENTS

RSF, WAA and AP-F would like to acknowledge support from the Brazilian agencies CNPq (478031/2013-0, 302923/2015-2 and 302880/2013-5, respectively) and FAPESP (2015/16191-5 and 2015/24018-1).

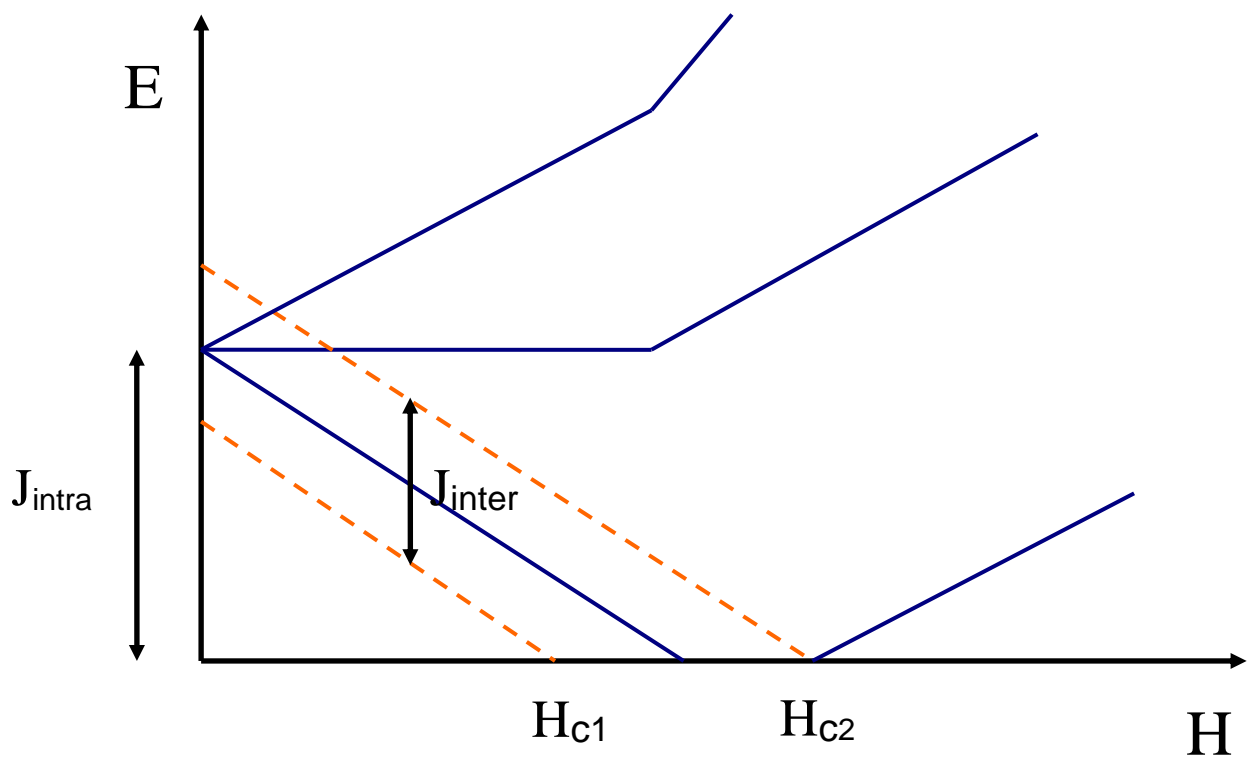

**Figure 1**. Spin levels scheme showing the zero-field gap due to the intradimer interactions. The levels are dispersed, forming bands due to the interdimer interactions that evolve in magnetic fields due to the Zeeman coupling.

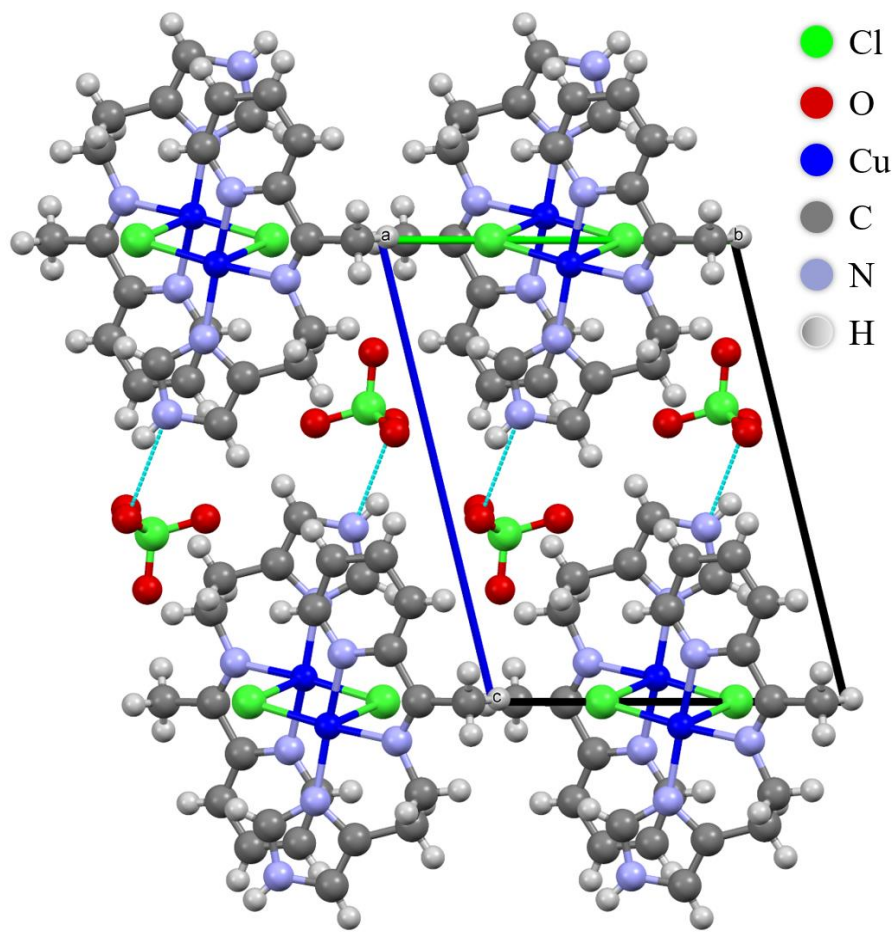

**Figure 2.** Perspective view of the crystallographically [Cu$_2$(apyhist)$_2$Cl$_2$](ClO$_4$)$_2$ unit with the atom labelling. The structure is consolidated by extensive intermolecular hydrogen bonds between binuclear species through the perchlorate ions, where each dimer interacts with six neighboring symmetry-related molecules.

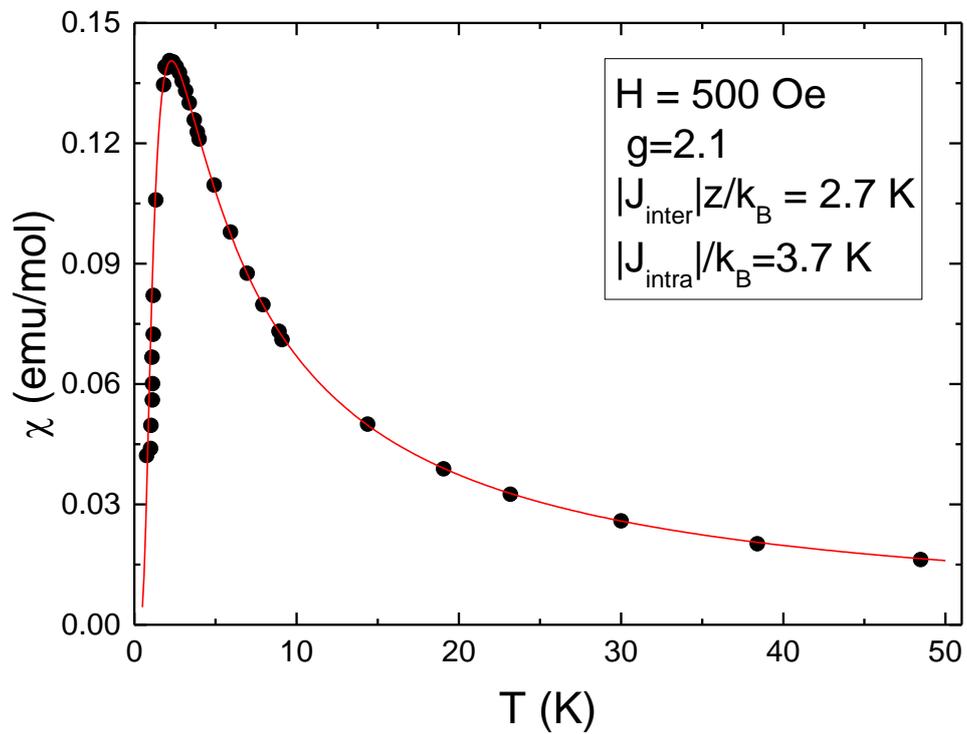

**Figure 3**. Temperature dependence of the magnetic susceptibility. The line represents the fit to the numerical calculations represented by equation 3.

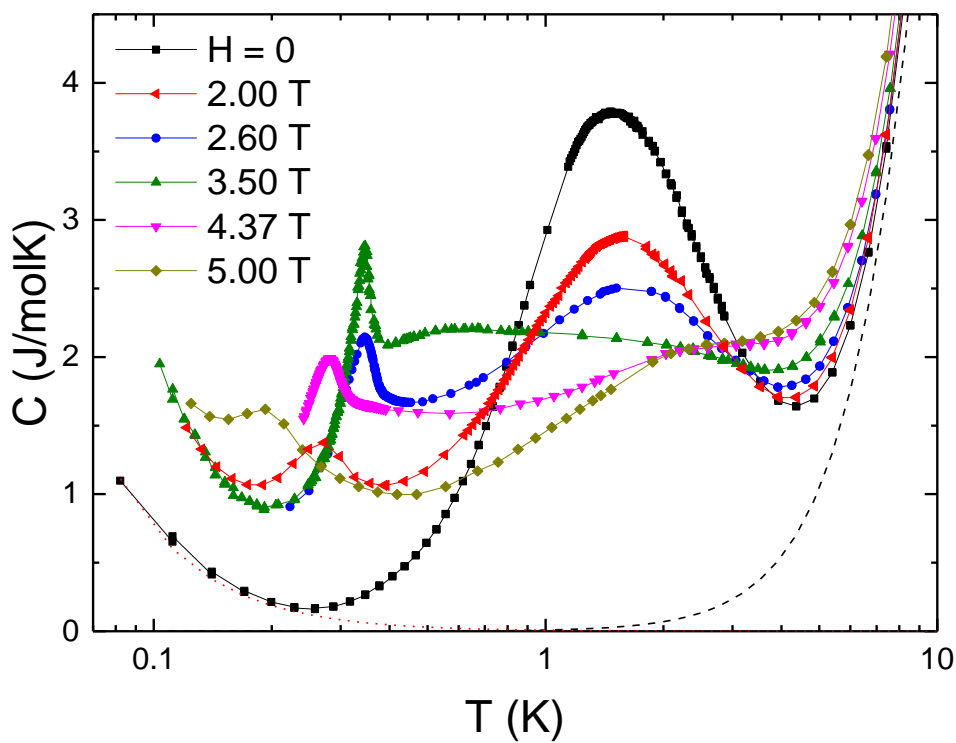

**Figure 4.** Specific heat as a function of the temperature measured under different applied magnetic fields. The dashed black and dotted red lines represent the lattice $C_l(T) = aT^3 + bT^5 + cT^7$ and nuclear $C_n(T) = \beta T^{-2}$ contributions with fitted coefficients $a$ = 2.9E-2, $b$ = -6.5E-5, $c$ = 6E-8 and $\beta$ = 7.5E-3. The continuous lines are only provided to serve as guides.

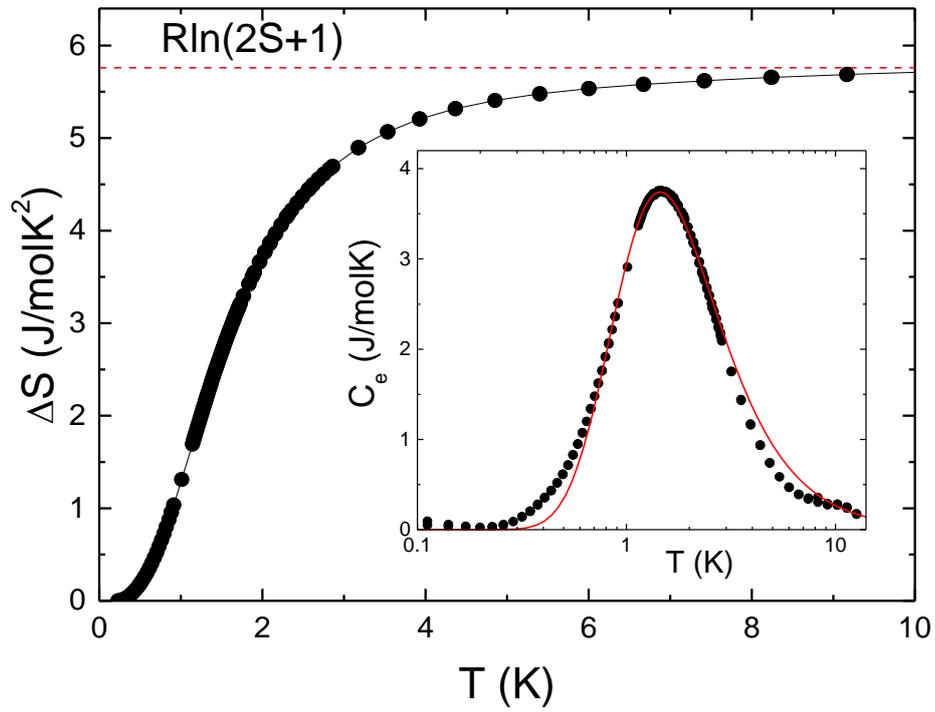

**Figure 5.** Integrated electronic magnetic entropy obtained after the subtraction of the lattice and nuclear contributions to the specific heat. The inset shows the electronic specific heat and the red continuous line represents the calculated specific heat for isotropic Heisenberg $S = ½$ dimers.

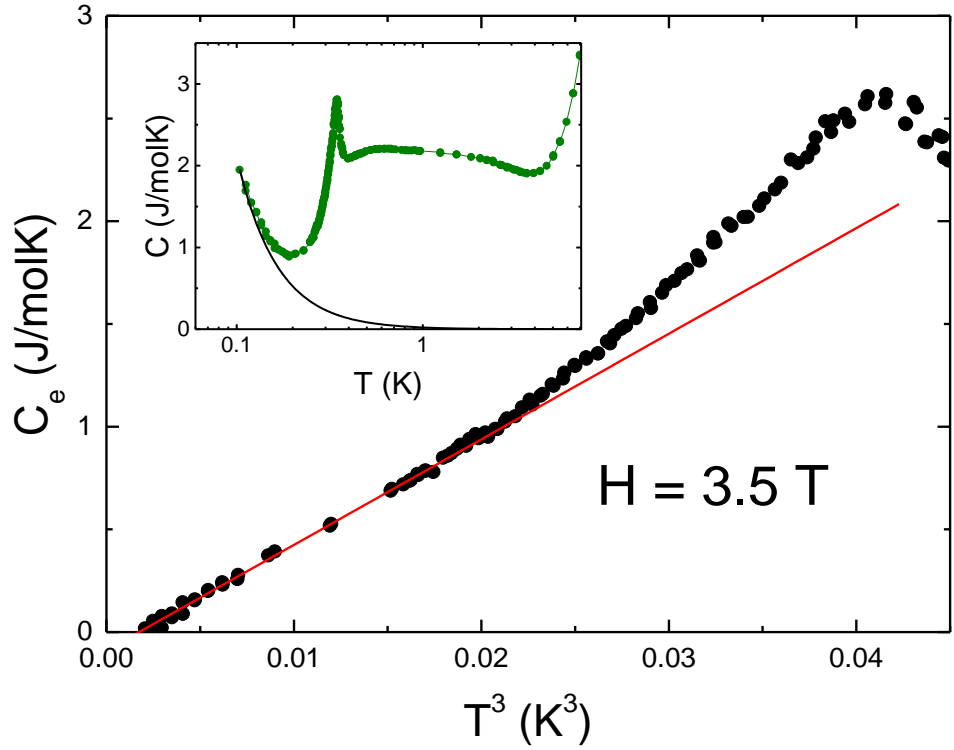

Figure **6.** The inset shows the total specific heat measured under H = 3.5T and the black line represents the nuclear contribution $C_n(T) = \beta T^{-2}$ with $\beta$ = 2.1E-2. The main panel shows the electronic specific heat, which shows the $T^3$ behavior characteristic of antiferromagnetic magnons.

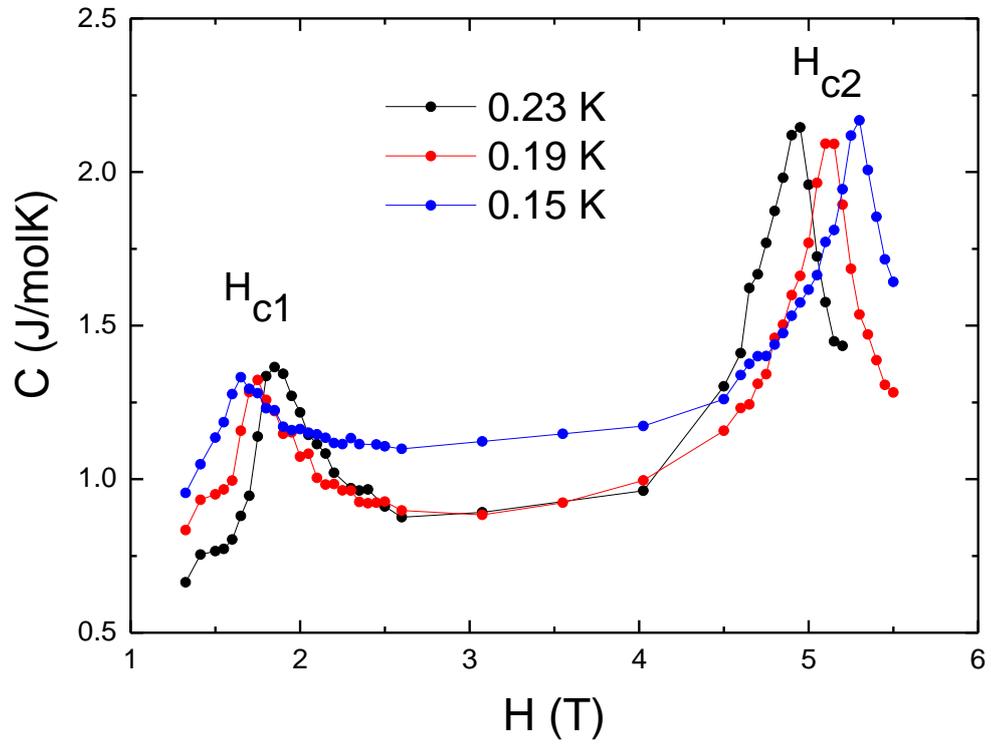

**Figure 7.** Magnetic field dependence of the specific heat measured at different temperatures as described in the text. The two peaks at $H_{c1}$ and $H_{c2}$ determine the boundary of the field-induced ordered phase.

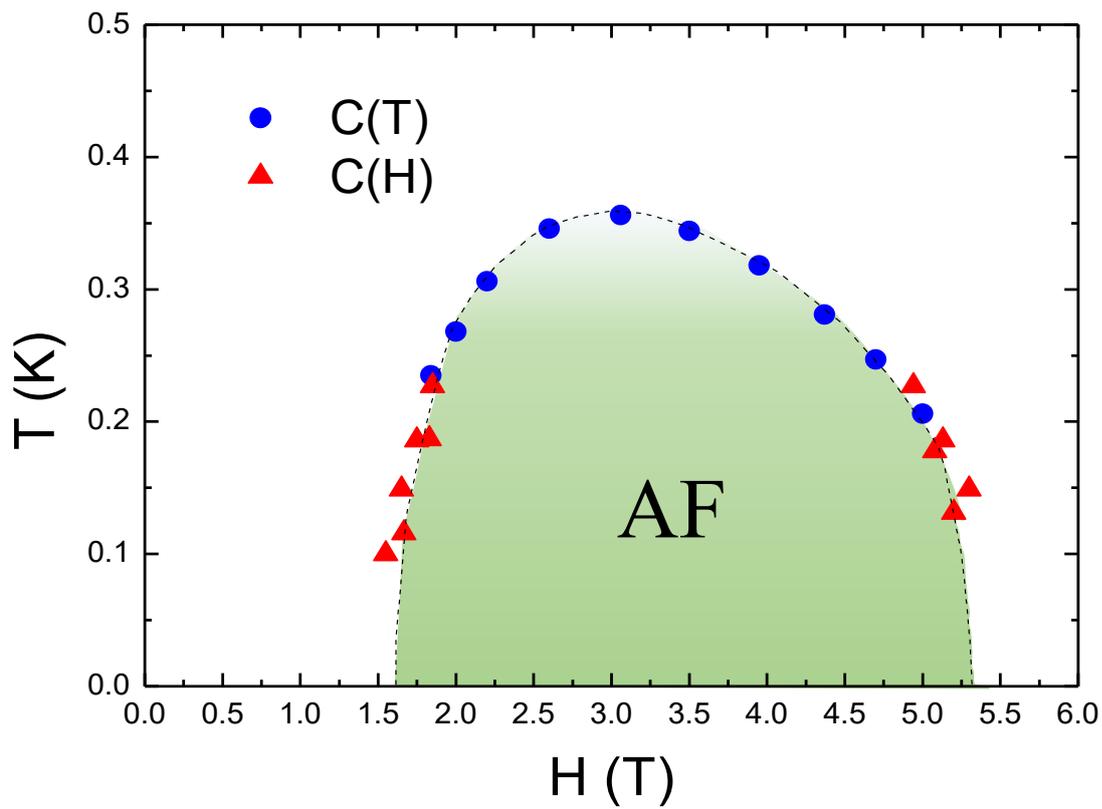

**Figure 8.** The magnetic phase diagram of [$Cu_2$(apyhist)$_2Cl_2$]($ClO_4$)$_2$ from the temperature and magnetic field scans of the specific heat showing the Field-induced long-range antiferromagnetic (AF) ordered phase.